\documentstyle[emulateapj,epsfig]{article}
\lefthead{Poli et al.}
\righthead{ }

\begin{document}
\title{THE EVOLUTION OF THE LUMINOSITY FUNCTION IN DEEP FIELDS:\\ A
COMPARISON WITH CDM MODELS}
\author{\sc F. Poli$^1$,
N. Menci$^1$, E. Giallongo$^1$, A. Fontana$^1$, S. Cristiani$^{2,3}$,
S. D'Odorico$^4$, 
}
\bigskip

\affil{$^1$ Osservatorio Astronomico di Roma, via dell'Osservatorio, I-00040
Monteporzio, Italy\\
\noindent
$^2$ ST European Coordinating Facility, Karl Schwarzschild Strasse 2,
D-85748 Garching, Germany\\
\noindent
$^3$ Dipartimento di Astronomia dell'Universit\`a di Padova, 
Vicolo dell'Osservatorio 2, I-35122 Padova, Italy\\
\noindent
$^4$ European Southern Observatory, Karl Schwarzschild Strasse 2,
D-85748 Garching, Germany\\
}
\begin{abstract}
The galaxy Luminosity Function (LF) has been estimated in the rest
frame B luminosity at $0<z<1.25$ and at $1700$ {\AA} for $2.5<z<4.5$
from deep multicolor surveys in the HDF-N, HDF-S, NTT-DF. The results
have been compared with a recent version of galaxy formation models in
the framework of hierarchical clustering in a flat Cold Dark Matter
Universe with cosmological constant. The results show a general
 greement for $z\lesssim 1$, although the model LF has a steeper
average slope at the faint end; at $z\sim 3$ such feature results in
an overprediction of the number of faint ($I_{AB}\sim 27$) galaxies,
while the agreement at the bright end becomes critically sensitive to
the details of dust absorption at such redshifts. The discrepancies
at the faint end show that a refined treatement of the physical
processes involving smaller galaxies is to be pursued in the
models, in terms of aggregation processes and/or stellar feedback
heavily affecting the luminosity of the low luminosity objects.  The
implications of our results on the evolution of the cosmological star
formation rate are discussed.
\end{abstract}


\section{INTRODUCTION}

The evolution of the luminosity function (LF) is a fundamental probe
for cosmological theories of galaxy formation. Indeed, the red and blue/UV
luminosities are related to the galaxy stellar mass and star formation
rate, respectively. Thus the evolution of the LF reflects the differential
contribution of different galaxy types to the cosmic history of mass growth
and of star formation which are the main outcomes of hierarchical models
for galaxy formation.

Adopting a Schechter fit to the LF, recent estimates (Zucca et
al. 1997, Marzke \& da Costa 1997, Folkes et al. 1999) point toward a
relatively steep power law at the faint end (with a power index
$\simeq 1.2$), although an excess of dwarf blue galaxies relative to
the Schechter fit is found at the faint end.

In the intermediate redshift range (up to $z\sim 1$), first steps
toward the evaluation of the LF evolution were undertaken by the
Canada France Redshift Survey (CFRS, Lilly et al. 1995) and by the
Autofib Redshift Survey (Ellis et al. 1996, Heyl et al. 1997). These
surveys have shown some increase in the number density of the fainter
population together with some increase in the luminosity of the
brighter blue fraction. These two effects are responsible for the
strong increase with $z$ of the average cosmic UV luminosity density (by
a factor 5--10) in the redshift interval $0<z<1$.

Deep multicolor surveys of galaxies represent an effective way to
explore the galaxy distribution in the redshift interval $1<z<5$.
Successful spectroscopic confirmation was obtained for the brightest
fraction by Steidel et al. (1996,1999). For the bulk of the population
reliable photometric redshifts are currently used (e.g. Connolly et
al. 1997; Giallongo et al. 1998, Fernandez-Soto, Lanzetta \& Yahil 1999;
Fontana et al. 2000).

The average UV luminosity density resulting from these studies shows a
flat distribution at $z>1$ that extends up to $z\sim 4.5$. Such a
result represents a strong constraint to hierarchical galaxy formation
theories that predict a significant contribution by a large
population of small objects. In a previous paper we showed that when
the effects of the sample magnitude limits are included in the models,
a steadly decline in the predicted UV luminosity density appears in
contrast with the observed behavior (Fontana et al. 1999).

Attempts to clarify the contribution of different (in mass and SFR)
galaxy populations to the global UV luminosity density, requires an
evaluation of the high redshift LF down to the faintest accessible
magnitudes. In this Letter we present a first estimate of the
intermediate/high-$z$ luminosity function using a composite deep
multicolor sample of about 1200 galaxies with reliable photometric
redshifts down to $I_{AB}=27.5$.  The depth of this data sample,
together with its relatively large area, allows a direct comparison of
the observed LF shape and evolution with theoretical predictions. Such
comparison is performed using a semianalytic implementation of recent
hierarchical models of galaxy formation.

\section{THE DATA SAMPLE}
The analyzed dataset covers a wavelength range from the UV to the K
bands, and observations were taken from ground based telescopes and
from the HST as well. The first field, known as New Technology
Telescope deep field (NTTDF) consists of an area of
$4.84\hspace{0.2cm}$ arcmin$^2$ where optical and near-IR UBVRIJK
observations have been taken at the ESO NTT telescope with various
instrumentation (SUSI, SUSI-2, SOFI).  Further details about the
multicolor catalog can be found in Arnouts et al.  (1999) and in Fontana
et al. (2000).

We have also used the Hubble Deep Fields North and South catalogs
provided by Fernandez-Soto et al. (1999), with an overall area of
$3.92$ arcmin$^2$ and 4.22 arcmin$^2$ respectively.  After appropriate
selections to remove contamination by stars and low signal-to-noise 
ratio (S/N) regions, we
applied our photometric redshift code to the data down to
$I_{AB}=27.5$. A detailed description of this procedure, along with
photometric $z$ catalogs, can be found in Fontana et al. (2000).

In addition, in the present evaluation of the luminosity function, we
have also considered one of the main systematic errors affecting the
estimates of the galaxy total absolute magnitudes at various $z$,
which is the surface brightness cosmological dimming effect. To
correct for this effect one has to recover the total galaxy emitted
flux. To estimate the systematic losses associated with any recovering
procedure, we performed simulations of synthetic galaxies (having
exponential intensity profiles with an ellipticity of 0.5) with
different apparent magnitudes, as seen with the appropriate S/N in the
HDF images. Then we compared the total input flux with the one
obtained by extrapolating the intensity profiles as computed in the
SEtractor package (Bertin
\& Arnouts 1996).  The simulations show that the difference $\Delta I$
between the input and the measured apparent magnitudes increases from
0.1 to 0.25 when the input magnitude increases from $I\sim 25$ to
$I\simeq 27.2$. For these reasons, although the catalogs in the HDFs
go deeper, we will confine the analysis of the LF at magnitudes $I\leq
27.2$ in the HDFs (and 25.7 in the NTTDF) where errors in the estimate
of the total magnitudes are small and do not affect our main results
on the shape of the LF.

\section{ESTIMATING THE LUMINOSITY FUNCTION}

Several methods are available in the literature (see Efstathiou, Ellis
\& Peterson 1988 for a discussion): here we choose to adopt the
classical $1/V_{max}$ estimator (Schmidt 1968) jointly with the
Sandage,Tammann \& Yahil (1979) maximum likelihood fit for a Schechter
function. In the $1/V_{max}$ method, for any given redshift bin
($z_1,z_2$) an effective maximum volume is assigned to each object.
This volume is enclosed between $z_1$ and $z_{up}$, the latter being
defined as the minimum between $z_2$ and the maximum redshift at which the
object could have been observed given the magnitude limit of the
sample.

Combining data from separate fields with different magnitude
limits, we then compute the galaxy number density $\phi(M_{B},z)$ in
every $(\Delta z,\Delta M_{B})$ bin as follows:

\begin{equation}
\phi(M_{B},z)=
\frac{1}{\Delta M_{B}}\sum_{i=1}^{N}\left[
\sum_j \omega(j)\int_{z_{1}}^{z_{up}(i,j)}
\frac{dV}{dz}dz
\right]^{-1}
\end{equation}

where $\omega(j)$ is the area in  units of steradians corresponding
to the field
$j$ and $N$ is the total number of objects in the chosen bin.  The
number of fields involved in the sum over index j is restricted to the
ones with a faint enough magnitude limit for the i-th object to be
detected.  In this way each galaxy has a different $z_{up}(j)$ in
every field and the overall volume available for this object is
obtained summing the corresponding $V_{max}(j)$.

On the other hand the STY technique, once assumed a Schechter
behaviour for the LF, is an endeavour to maximize the likelihood of
representing the observed set of galaxies with the best fit parameters
of the Schechter function. Assuming that, in an appropriately
thin bin of redshift, the number of sources with redshift between $z$
and $z+dz$, and with absolute magnitude between $M_B$ and $M_B+dM_B$
can be factorized as $\phi(M_B,z)dzdM_B=\rho(z)\psi(M_B)dzd\,M_B$, where
$\rho(z)$ is the density of galaxies at redshift $z$, that can be
considered constant within the bin. The function $\psi(M_B)$ is
taken to have a Schechter form, parametrized by the characteristic
absolute magnitude $M_B^*$, by the logarithmic slope at the faint end
$\alpha$ and by the normalization $\psi_*$.

If we choose an appropriate redshift bin $(z_1,z_2)$, it is possible
to give an estimation of $\alpha$ and $M_{B}^{*}$ considering the
probability density $p_{i,j}$ to find a galaxy with redshift in the
range $z_i,z_i+dz$ and absolute magnitude between $M_i$ and $M_i+dM$
in the j-th magnitude-limited field and maximizing the likelihood
of observing the set of galaxies that comes from the
surveys, wich is simply given by the product of the single
probabilities:
\begin{equation}
\prod_j\prod_{i=1}^{N_j}p_{i,j}=\prod_j\prod_{i=1}^{N_j}\frac
{\rho(z_i)\psi(M_i)dzdM}{\omega(j)\int_{z_1}^{z_2}\rho(z)\frac{dV}{dz}dz
\int_{-\infty}^ { M_{lim}^{i,j}(z) }\psi(M)dM}
\end{equation}
$\omega(j)$ beeing the field area in steradians, and $M_{lim}^{i,j}(z)$
the absolute magnitude value that the i-th object, if detected with the
magnitude limit $m^{j}_{lim}$ in the j-th survey, should have
at that redshift. This value clearly depends on the details of the spectrum
of each object. Here j runs over the number of fields where the i-th
galaxy can be detected, each one containing $N_j$ objects.

The value of $\phi^{*}$ is then obtained summing the density in the $z,M$
space for
every galaxy, taking account of all the fields where it could be
detected (i.e. the fields with enough bright magnitude limit):
\begin{equation}
\phi^{*}=\sum_{i=1}^{N_j}\left[
\sum_{j detect}\omega(j)\int_{z_1}^{z_2}\frac{dV}{dz}dz
\int_{-\infty}^{M_{lim}^{i,j}(z)}\psi(M)dM\right]^{-1}
\end{equation}
As for an appropriate selection in magnitude of the sample, it is
important to bear in mind that once a rest frame wavelength $\lambda$
is chosen (e.g. the $4400$ \AA \hspace{0.2cm} band), this corresponds
to different observed wavelengths $\lambda(1+z)$ when $z$ runs inside
the redshift bin $(z_1,z_2)$.  Since our aim is to choose
appropriately a complete subsample selected in the rest-frame, we
ought to take into account the different redshifts of each galaxy and
the details of the spectra as well, to be sure that we are selecting
objects in a coherent way. One major concern is fixing the bins in
redshift in a suitable way for matching the observed bands once the
wavelength displacement is taken into account.  Subsequentely the
right selection criterion can be achieved by comparing the photometric
limit with the $\lambda(1+z)$ magnitude value taken from the
spectrum. This implies a limiting magnitude $\leq 27.2$ at a
wavelength of $4400(1+z)$ {\AA} for $z\leq 1.25$ and $\leq
27.2$ at $1700(1+z)$ {\AA} for $z>2.5$ in the HDFs.

\begin{figure*}
\epsfig{file=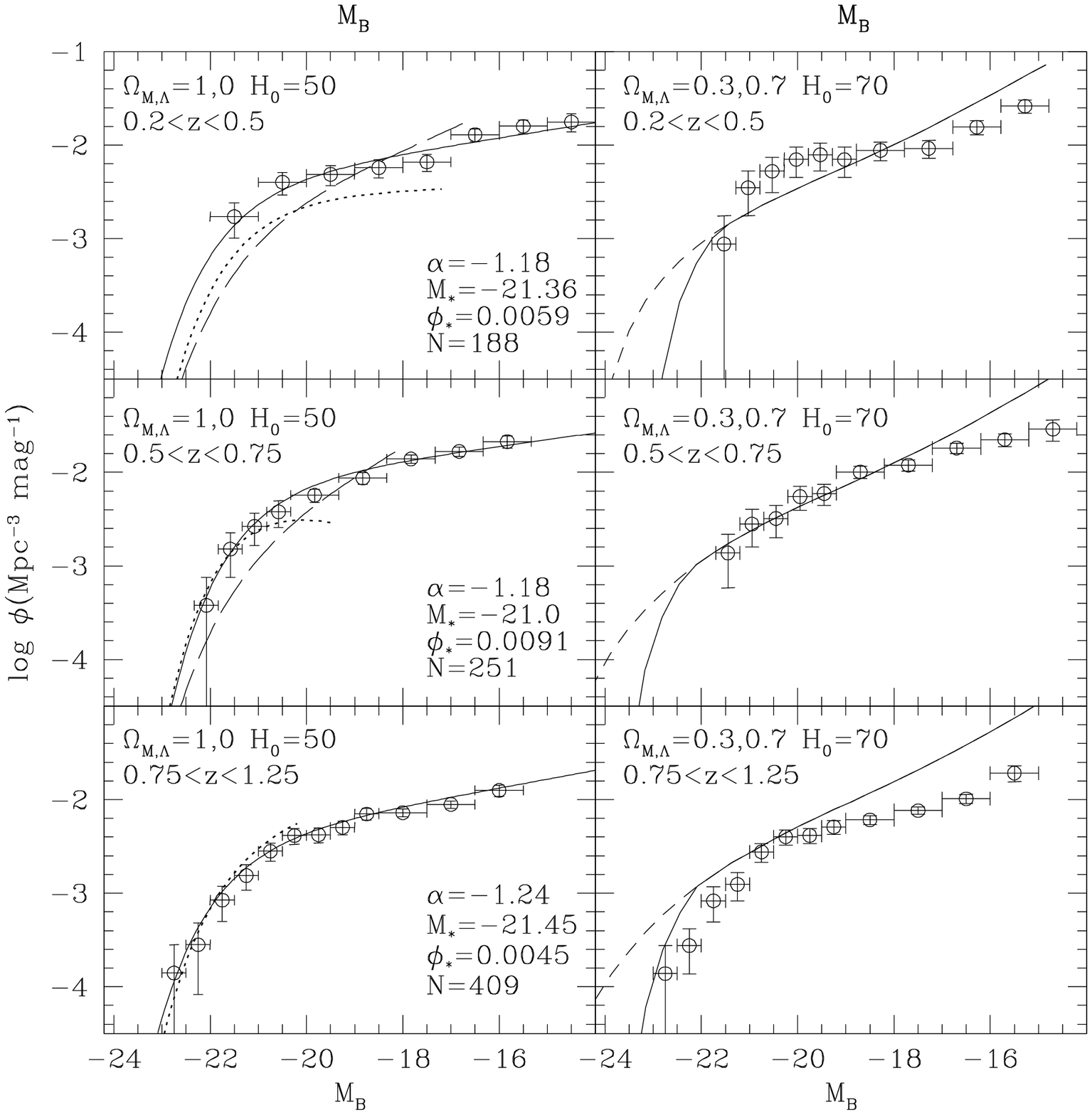,height=18truecm}
{\footnotesize
\noindent 
Fig. 1.-Rest frame B luminosity function of field galaxies in
different redshift bins and for two cosmological models.  {\it Left
panel:} The continuous curves are the Schechter LFs resulting from the
Maximum likelihood fit to our composite galaxy sample. Dotted curves
are the Schecter functions derived from the CFRS survey (Lilly et
al. 1995), while the dashed curves are derived from the Autofib
redshift survey (Heyl et al. 1997). {\it Right Panels:} Continuous
curves are the CDM model predictions discussed in the text, including
dust absorption with a SMC extinction curve (different extinction curves
used in Fig.2 do not produce appreciable changes in the B band LFs).
Dashed curves are the CDM dust-free model predictions.
}
\end{figure*}

\section{THE COSMOLOGICAL EVOLUTION OF THE LF}

The luminosity function at the rest frame B magnitude in the AB system,
$M_B$, is shown in Fig.~1 for three redshift bins
($0.2-0.5,0.5-0.75,0.75-1.25$). The left panels refer to a critical
universe with $\Omega_M=1$, $\Omega_{\Lambda}=0$ and $H_0=50$ km
s$^{-1}$ Mpc$^{-1}$, while the right panels refer to a flat universe
dominated by the cosmological constant $\Omega_M=0.3$,
$\Omega_{\Lambda}=0.7$ and $H_0=70$ km s$^{-1}$ Mpc$^{-1}$. The rest
frame B magnitude is computed from the best fit theoretical spectral
energy distribution used to derive the photometric redshift.  The
magnitude limit of the sample has been evaluated at the same
rest-frame wavelength centered in the rest frame B band, which roughly
corresponds to the observed V,R,I bands for the three redshift bins,
respectively.  In addition we show in Fig.2 the $1700${\AA} rest-frame
LF in the bins $2.5<z<3.5$ and $3.5<z<4.5$ for the same cosmologies.

The best fit values of the Schechter parameters obtained from the
maximum likelihood method are summarized in table 1 together with the
relevant sample parameters for both the considered cosmologies. In the
highest redshift bin ($3.5<z<4.5$) the small number of objects
prevented an accurate estimate of the Schechter parameters. The
Schechter curves corresponding to the tabulated fitting parameters are
shown in Figs. 1,2 only for the case of the critical universe (left
panels) for comparison with previous LFs computed in the same cosmology. In
particular, the Schechter LFs from the Canada France Redshift Survey
and the Autofib survey are also shown for $z<1.25$ as dotted and
dashed curves, respectively.  An overall agreement with the
spectroscopic data is evident at bright magnitudes, supporting the
reliability of our LF photometric estimation. The variance between
spectroscopic surveys could be due to different k-corrections applied
for the estimate of the rest-frame blue 
magnitudes from samples selected in different bands.
As discussed in Sawicki et al. (1997), the uncertainties in the
photometric redshift estimates do not affect appreciably the
behavior of the LF: suitable Monte Carlo simulations have shown that
perturbations arising from redshift uncertanties result in effects
smaller than one sigma error bars in the $1/V_{max}$ estimator and in
small changes in the best-fit Schechter parameters.  In our case the
good match with CFRS leaves little space for this kind of substantial
changes in the distribution.

It can be seen that from $z\sim 0.2$ up to $z\sim 1.25$, there is no
evidence of a significant trend with redshift in the faint-end slope
parameter $\alpha$, which remains close to its local value $\alpha
\simeq -1.2$ (Zucca et al. 1997). Also, the characteristic magnitude
$M_*$ shows a mild brightening with the look-back time.At high redshifts 
a steepening effect is evident in the $1700${\AA} LF
(Fig.2, left panel), where the slope parameter reaches a value of $-1.37$,
consistent within uncertainties with the $\alpha=-1.6$ found by
Steidel et al. (1999) in an analogous redshift bin. We note however that
in this high redshift bin the slope is weakly constrained by the
present depth of the galaxy sample.

\begin{figure*}
\epsfig{file=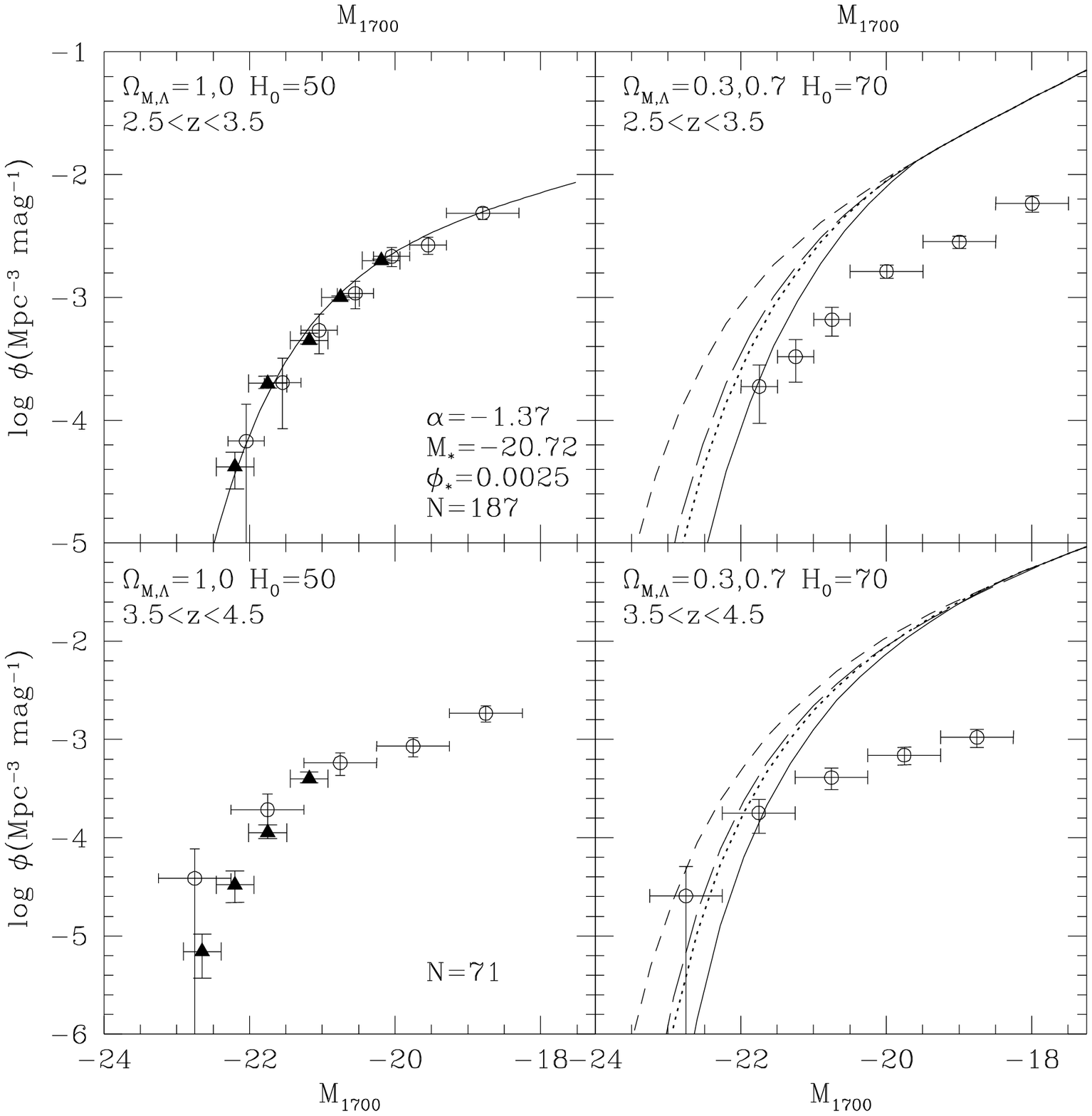,width=18truecm}
{\footnotesize\noindent
Fig. 2.-Rest frame ultraviolet luminosity function at $z\sim 3$ and
$z\sim 4$.  Filled symbols are derived from the spectroscopic survey
by Steidel et al. (1999).  {\it Left Panel:} Continuous curve is
derived from the Schechter Maximum Likelihood fit as in Fig.1. {\it
Right Panel:} Symbols as in the corresponding panel in Fig.1 except
for the dotted or long-dashed curves which refer to the Milky Way and
Calzetti extinction curves, respectively.
}
\end{figure*}
\vspace{0.4cm}

\section{A COMPARISON WITH HIERARCHICAL MODELS FOR GALAXY EVOLUTION}

We compare our data with our rendition of the semianalytic models. The
structure of such rendition is described in Poli et al. (1999). We
updated our model to implement some of the most recent improvements
inserted in the recent versions of semianalytic models, following the
lines of Cole et al. (2000). In particular we now adopt the Lacey \&
Cole (1993) dynamical friction timescale, the new star formation and
feedback recipes and the new modelization for hot gas distribution
implemented by the above authors. This allows us to obtain a Tully-Fisher
relation in reasonable agreement with observations together with a
good fit to the local luminosity function, the two observables that
have been used to calibrate the free parameters governing the star
formation and the feedback processes. In addition we have included
dust absorption, which is modelled as in Poli et al. (1999).
We have checked that the predictions of our model agree with those from
Cole et al. (2000) for both the local observables and the global cosmic
star formation history.

\vspace{0.4cm}
\centerline {Table 1. }

\centerline{Parameters of the Schechter function fits}
\begin{tabular}{lcccc}
\hline
\hline
z range   &$\alpha ^a$&$M_* ^b$&$\phi_*$&N \\
\hline
02-0.5    &$-1.18\pm 0.05$ &$-21.36\pm0.37$ &0.0059 &188 \\
 ~         &$-1.19\pm 0.05$ &$-21.03\pm0.38$ &0.0086 & ~  \\
0.5-0.75  &$-1.18\pm 0.06$ &$-21.00\pm0.29$ &0.0091 &251 \\
 ~         &$-1.19\pm 0.06$ &$-20.75\pm0.30$ &0.01   & ~  \\
0.75-1.25 &$-1.24\pm 0.06$ &$-21.45\pm0.24$ &0.0045 &409\\
 ~         &$-1.25\pm 0.05$ &$-21.38\pm0.25$ &0.0041 & ~  \\
2.5-3.5   &$-1.37\pm 0.20$ &$-20.72\pm0.37$ &0.0025 &187 \\
 ~         &$-1.37\pm 0.19$ &$-20.84\pm0.37$ &0.0023 & ~  \\
\hline
\end{tabular}

$^a$ The second row for each $z$ bin refers to the $\Omega_{M}=0.3$,
$\Omega_{\Lambda}=0.7$ cosmology.

$^b$ $M_*$ is $M_B$ (AB) except in the $z=2.5-3.5$ bin where it is computed
at
1700 {\AA}, $M_{1700}$

\vspace{0.4cm}

To compare with the data we focus on the flat $\Lambda$CDM (Cold Dark Matter)
cosmology.
Such choice is favoured by a significant amount of recent
observational results, from the high-fraction of the baryons-to-DM
ratio in galaxy clusters (see, e.g., White \& Fabian 1995) to the
high-redshift SNe (see Perlmutter et al. 1999). When inserted into
galaxy formation models, the above set of parameters also yields a
cosmic star formation history in reasonable agreement with
observations (see Fontana et al. 1999), in contrast, e.g., with the
Standard CDM with $\Omega_M=1$ and $\Omega _{\Lambda}=0$.

Fig.1 and Fig. 2 (right panels) show a comparison of the observed
luminosity functions derived from our composite sample and our CDM
model. First we note a general agreement for $M_{B}<-20$ and
$z\lesssim 1$. Both data and predictions show weak luminosity
evolution resulting from a balance between a mild decrease with $z$ of
the number of massive objects and the increase of their star formation
activity and hence of their blue luminosity. At the same time, the
normalization of the LF rises by a factor of $\sim 2$.

At low luminosities the theoretical LF appears steeper than the
observed one, with an excess at the faint end that becomes larger
with increasing redshift.  In addition, at $z>3.5$ the model tends to
slightly underestimate the magnitudes of bright galaxies when the
model with larger dust extinction ($\sim 1$ mag, in agreement with
Pettini et al. 1998) is considered.  Such results show that a
refinement of the models is required, both in terms of dynamical
processes (like merging of satellites in common halos, see Somerville, 
Primack, \& Faber 2001) and in terms of
stellar processes (like the feedback from Supernovae, appreciably
affecting the LF at the faint-end). Indeed, the apparent agreement of
the predicted vs. observed UV luminosity density at $z>3$ (Fontana
et al. 1999) results from the balance in the LF between the excess of
the predicted dwarfs and the deficit of predicted bright galaxies, as
shown in the right panels of fig. 2.

This shows that the $z$-resolved LFs constitute a particulary powerful 
way of constraining the two most uncertain processes in the 
theoretical modelling, i.e., the dust absorption affecting the prediction of  
the number of bright galaxies and the Supernovae feedback affecting  
the slope of the predicted LF at the faint end. 

The above considerations suggest that the comparison of the observed
and predicted UV luminosity density at $z\geq 3$ is critical at the
bright end. Indeed, the details of the dust extinction are important
at the bright end of the steep LF function since the dust extinction
affects not only the observed UV luminosities but also the number of
galaxies that are detected within the magnitude limit. Some spectroscopic
information for the brightest sources can better constrain the dust
content at these high $z$.

For such a reason, the comparison between the predicted and observed UV
luminosity densities should be performed over wider areas of the sky,
to reduce fluctuations in the number of bright sources, and using
different limiting magnitudes, extended to the faintest limits, to
explore the dust content as a function of the galaxy star formation
rate and mass.  This will allow us to sample the shape of the LF at the
faint end, which is crucial to assess whether complementary physical
processes have to be included in the current hierarchical models.


\begin{references}

\reference{}{Arnouts, S., D'Odorico, S., Cristiani, S., Zaggia, S.,
Fontana, A., Giallongo, E., 1999, A\&A, 341, 641}

\reference{}{Bertin, E., \& Arnouts, S., 1996, A\&AS, 117, 393}

\reference{}{Cole, S., Lacey, C.G., Baugh, C.M.; Frenk, C.S., 2000,
MNRAS, 319, 168}

\reference{}{Connolly, A. J., Szalay, A. S., Dickinson, M., SubbaRao, M. U.,
Brunner, R. J., 1997, ApJ, 486, L11}

\reference{}{Efstathiou, G., Ellis, R. S., \& Peterson, B. A., 1988,
MNRAS, 232, 431}

\reference{}{Ellis, R.S., Colless, M.M., Broadhurst, T.J., Heyl, J.S.,
Glazebrook, K. 1996, MNRAS, 280, 235}

\reference{}{Fernandez-Soto, A., Lanzetta, K. M., Yahil, A., 1999,
ApJ, 513, 34}

\reference{}{Folkes, S., et al., 1999, MNRAS, 308, 459}

\reference{}{Fontana, A., D'Odorico, S., Poli, F., Giallongo, E.,
Arnouts, A., Cristiani, S., Moorwood, A., Saracco, P., 2000, AJ, 120,
2206}

\reference{}{Fontana, A., Menci, N., D'Odorico, S., Giallongo, E., Poli, F.,
Cristiani, S., Moorwood, A., \& Saracco, P., 1999, 310, L27}

\reference {}{Giallongo, E., D'Odorico, S., Fontana, A., Cristiani, S.,
Egami, E., Hu, E., McMahon, R. G.  1998, AJ, 115, 2169}

\reference{}{Heyl, J., Colless, M., Ellis, R.S., Broadhurst, T. 1997,
MNRAS, 285, 613}

\reference{}{Lacey, C., \& Cole, S., 1993, MNRAS, 262, 627}

\reference{}{Lilly, S., Tresse, L., Hammer, F., Crampton, D., Le F\`evre,
O.,
1995, ApJ, 455, 108}

\reference{}{Lin, H., Yee, H.K.C., Carlberg, R.C., Morris, S.L., Sawicki,
M.,
Patton, D.R., Wirth, G., Shepherd, C.W. 1999, ApJ, 518, 533}

\reference{}{Marzke, L.O. \& da Costa, L.N. 1997 AJ, 113, 185}


\reference{}{Perlmutter, S. et al. 1999, ApJ, 517, 565}

\reference{}{Pettini, M., Kellogg, M., Steidel, C. C.,  Dickinson, M.,
Adelberger, K. L., Giavalisco, M.,  1998, ApJ, 508, 539}

\reference{}{Poli, F., Giallongo, E., Menci, N., D'Odorico, S., \& Fontana,
A.
 1999, ApJ, 527, 662}

\reference{}{Sandage, A., Tammann, G. A.,  \& Yahil, A., 1979, ApJ, 232,
352}

\reference{}{Sawicki, M.J. Lin, H., Yee, H.K.C. 1997, AJ, 113, 1}

\reference{}{Schmidt, M.  1968, ApJ, 151, 343}

\reference{}{Somerville, R. S., Primack, J. R., Faber, S. M. 2001, MNRAS, 320,
504}

\reference{}{Steidel, C.C., Adelberger, K.L., Giavalisco, M., Dickinson, M.
 Pettini, M. 1999, ApJ, 519, 1}

\reference {}{Steidel, C. C., Giavalisco, M., Pettini, M., Dickinson, M.,
\& Adelberger, K. L. 1996, ApJ, 462, L17}

\reference{}{White, D. A., \& Fabian, A. C.,  1995, MNRAS, 273, 72}

\reference{}{Zucca, E., et al. 1997, A\&A,326, 477}




\end{references}
\end{document}